# Title: Electron confinement within a fluctuating "box" in liquid water

**Authors:** Korenobu Matsuzaki[1,2], Hikaru Kuramochi[1,2]†, Tahei Tahara[1,2]*

**Affiliations:**

[1]Molecular Spectroscopy Laboratory, RIKEN; Wako, 351-0198, Japan.

[2]Ultrafast Spectroscopy Research Team, RIKEN Center for Advanced Photonics (RAP), RIKEN; Wako, 351-0198, Japan.

*Corresponding author. Email: tahei@riken.jp

†Present address: Department of Materials Engineering Science, Graduate School of Engineering Science, The University of Osaka; Toyonaka, 560-8531, Japan.

**Abstract:** Electron confinement within a small volume is intriguing as a realization of the particle-in-a-box system, which appears in every quantum mechanics textbook. While the electron confinement is readily imaginable in solid-state systems, it also occurs in liquids, where the local voids in the liquid serve as confining "boxes." Confinement within these flexible cavities in liquids is expected to differ fundamentally from that in solids. Here, we experimentally investigate the electrons confined in liquid water, which are called hydrated electrons, using transient two-dimensional electronic spectroscopy. Our experiment reveals the large nonuniformity of the shape and the size of hydrated electrons with significant fluctuation at the timescale shorter than 30 fs.

**One-Sentence Summary:** Ultrafast spectroscopy reveals the elusive nature of electron confinement in liquid water.

**Main Text:**

Electrons are ubiquitous in our lives. Electricity results from the flow of electrons through metallic wires, and the reflection of light by a metallic mirror originates from the oscillation of electrons inside the metal caused by the incident light. While the electrons in motion are utilized in these two cases, it is also possible to confine electrons within a small volume defined by nanostructures, with a quantum dot (*1*) being a representative example. Such a system is intriguing from the viewpoint of fundamental quantum science because it can be treated as a real-world particle-in-a-box (*2, 3*). Indeed, the relationship between the quantum dot size and its emission wavelength can be rationalized based on the particle-in-a-box picture, i.e., a bigger box results in a smaller energy spacing and hence longer optical transition wavelength (*1*).

Such electron confinement is possible not only in solid-state systems but also in the liquid phase. Liquid is a flexible system whose molecular-scale structure constantly fluctuates. This fluctuation results in the occasional generation of local voids that are depleted of molecules, which serve as the "boxes" that confine electrons in the liquids. In sharp contrast to solid-state systems, where the confining "box" has a fixed shape, the shape of the "box" in liquid systems continuously fluctuates due to the movement of molecules. Therefore, the properties of the confined electrons in liquid, which are determined by the size and the shape of the voids, also fluctuate, making their full characterization nontrivial.

A representative example of electron confinement in a liquid can be found in liquid water (*4*), where an electron is trapped inside a cavity made of approximately four water molecules (*5*) (Fig. 1). This species is called a hydrated electron, which has attracted a lot of attention not only due to the fundamental academic importance but also because it plays major roles in radiation damages (*6*). Hydrated electrons have been intensively studied using absorption spectroscopy, yielding a characteristic broad absorption spectrum from visible to near-infrared wavelength (*7, 8*). The absorption spectrum is determined by the electronic energy levels of the hydrated electrons, which in turn are dictated by the shape and size of the cavity that confines the electron.

We can imagine that the shape of this "box" made of water molecules is approximately spherical, because the environment inside bulk liquid water is isotropic. Hydrated electrons then have the same symmetry as a hydrogen atom, and their electronic structure can also be theoretically described in a similar manner: the lowest four energy levels are denoted as an $s$-state and three $p$-states ($p_x$, $p_y$, and $p_z$), and the three $p \leftarrow s$ transitions have transition dipole moments that are perpendicular to each other. The major part of the observed absorption spectrum is attributed to the transition from the $s$-state to the three $p$-states (*9*), while the short wavelength tail of the absorption spectrum is attributed to the transitions to continuum (*9*) or to quasi-continuum states (*10*). For a perfectly spherical system, the three $p$-states are degenerate, and the three $p \leftarrow s$ transition frequencies are expected to be identical. In the meantime, when hydrated electrons are not perfectly spherical because the cavity made by water molecules has an asymmetric shape, the degeneracy of the three $p$-states is lifted. In fact, it is theoretically proposed that the three $p \leftarrow s$ absorption lines are centered at slightly different wavelengths due to the asymmetric shape of the cavity and that the broad absorption spectrum of hydrated electrons originates from the superposition of the three $p \leftarrow s$ absorption lines with a relatively narrow bandwidth at slightly shifted central wavelengths (*9*).

Schwartz *et al.* have proposed that the aforementioned theoretically deduced picture of hydrated electrons can be experimentally validated based on polarized hole-burning measurements (*11*). In this measurement, a narrowband pump is irradiated onto hydrated electrons to induce one of the three $p \leftarrow s$ transitions, e.g., the $p_x \leftarrow s$ transition. Then, when the system is probed with the same polarization as the pump, a depletion of the spectral intensity at the $p_x \leftarrow s$ transition frequency is expected because the system is already excited to the $p_x$ state by the preceding narrowband pump. This is called the hole-burning effect. If the probe is polarized perpendicular to the pump, on the other hand, the hole-burning effect is expected to be observed for the $p_y \leftarrow s$ and $p_z \leftarrow s$ transitions, i.e., at a frequency different from the pump, because the transition dipole moments of the three $p \leftarrow s$ transitions are orthogonal to each other. Schwartz *et al.* referred to this effect in their theoretical study as the "replica hole" effect (*11*). In both cases, the hole is expected to be observable only immediately after the pump irradiation when the shape and orientation of the hydrated electrons remain unchanged. As the pump-probe delay increases, the hydrated electrons change their shape or orientation, and the memory of the pump frequency in the system rapidly fades, resulting in the broadening and disappearance of the spectral hole.

From the experimental side, polarized hole-burning measurements have been performed on the hydrated electron system by several experimental groups. The Barbara group performed a pioneering experiment with the time resolution of 300 fs (*12*), and their experimental data supported the existence of the hole-burning signal that was theoretically predicted by Schwartz *et al.* (*11*). Their result, however, was not reproduced in the measurements performed with a time resolution of 100-200 fs by the groups of Laubereau and Schwartz, which showed no sign of hole burning (*13, 14*). The Wiersma group also performed the measurement with a much higher time resolution using 5-fs pulses and recorded the transient spectra at the pump-probe delay as short as 20 fs. Neither in their measurement was the hole burning observed (*15*). Thus, to the best of our knowledge, no convincing experimental evidence has been obtained so far for the hole burning of the absorption spectrum of hydrated electrons. At the same time, we notice that no group has studied the spectral response of the hydrated electron system at the pump-probe delay shorter than 20 fs with sufficient time resolution.

The time resolution in hole-burning measurements is limited by the Fourier transform relationship between the frequency bandwidth and the temporal duration of the narrowband pump, which means that the time resolution needs to be compromised to realize a high spectral resolution necessary for hole-burning measurements. Two-dimensional electronic spectroscopy (2DES) allows us to resolve this dilemma and realizes high time and spectral resolution simultaneously (*16*). In this method, instead of a narrowband pump beam, a pair of broadband ultrashort pulses is used as the pump. By performing the pump-probe measurements while varying the temporal separation between the paired pulses, we obtain a two-dimensional spectrum where the vertical and horizontal axes represent the pump and probe frequencies, respectively. The spectrum visualizes the correlation between the transition frequencies of the sample at the moments of pump and probe irradiation, which provides unique information about the system under interrogation. For example, it shows the flow of energy introduced into the system by the pump, which plays a central role in biological systems such as light-harvesting complexes or in artificial systems such as solar cells (*17-21*). Furthermore, it also contains spectroscopic information equivalent to the hole-burning measurement. More specifically, the 1D slice of a 2DES spectrum at a certain pump frequency corresponds to a hole-burning spectrum measured at that pump frequency. In this sense, a 2DES measurement can be regarded as a hole-burning measurement performed simultaneously at various pump frequencies, with the important additional advantage of the simultaneous realization of high spectral and time resolutions. In 2DES spectra, the hole-burning effect is observed as a diagonal elongation of a two-dimensional peak because the hole-burning signal appears at the position where the pump and probe frequencies are the same. Such a diagonally elongated signal indicates that the absorption spectrum of the sample arises from a superposition of multiple spectral components at each frequency. In this case, the absorption spectrum is said to be inhomogeneously broadened. On the other hand, when the diagonal elongation is absent, the spectrum consists of a single spectral component, and it is said to be homogeneously broadened. In this way, by examining the presence or absence of the diagonal elongation of a two-dimensional peak in a 2DES spectrum, it is possible to clarify if an absorption spectrum is homogeneously or inhomogeneously broadened (*22*).

The aim of this study is to perform hole-burning measurements on hydrated electrons with unprecedented temporal resolution based on 2DES and to elucidate the nature of electron confinement inside a flexible "box" in liquid water. Since hydrated electrons are short-lived transient species, such measurements require the extension of 2DES to "transient" 2DES (tr-2DES). In this technique, short-lived transient species are generated by the actinic pump irradiation prior to 2DES measurements. Subsequently, a pump pulse pair with varying temporal separation is irradiated to photo-excite the transient species, and the dynamics following the photo-excitation is interrogated using a probe pulse (Fig. 1). By using sub-10-fs pulses for the pump pulse pair and the probe pulse, we realized a measurement corresponding to transient hole-burning measurements with a high time resolution of 12 fs.

Hydrated electrons in our tr-2DES measurements were prepared by irradiating an intense deep ultraviolet actinic pump pulse at 257 nm onto a thin film of liquid water, where electrons were generated due to the two-photon ionization of water molecules. Initially, the electrons occupy the conduction band and are delocalized. Subsequently, those electrons become localized and solvated by several water molecules, forming hydrated electrons (*23*).

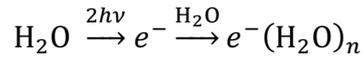

$$H_2O \xrightarrow{2h\nu} e^- \xrightarrow{H_2O} e^-(H_2O)_n$$

The hydrated electrons generated in this manner undergo relaxation from the excited state (*p*-state) to the ground state (*s*-state) on the timescale of 60 fs (*24*), which is followed by a further solvation process within 1 ps to yield fully equilibrated hydrated electrons (*25*). Those fully equilibrated hydrated electrons were used as the sample for the tr-2DES measurements in this study.

Fig. 2A shows the transient absorption spectrum of hydrated electrons measured 10 ps after the electron ejection by the actinic pump irradiation. It shows very broad absorption centered at ~740 nm, which agrees well with the transient spectrum of fully equilibrated hydrated electrons in the literature (*7, 8*). The good agreement of the spectra confirms that fully equilibrated hydrated electrons are successfully prepared in our experiment.

The properties of these fully equilibrated hydrated electrons were examined by applying tr-2DES measurements to them. For the first measurement, the temporal separation between the probe pulse and the pump pulse pair (the second pulse in the pair), i.e., the waiting time ($T_w$, Fig. 1), was 0 fs, and the polarization of the pump pulse pair was parallel to that of the probe pulse. The obtained tr-2DES spectrum in Fig. 2B shows a negative signal (indicated in red) that is diagonally elongated. The negative sign of the signal suggests that what we observe here is the bleach of the absorption upon the pump pulse pair irradiation, which is caused by the depletion of the hydrated electrons in the ground state. The diagonal elongation of the signal further indicates that the bandwidth of the bleach signal is narrower than the transient absorption spectrum of hydrated electrons (Fig. 2A) and that the bleach signal appears at the position corresponding to the pump frequency. In other words, the tr-2DES spectrum shows hole burning at each pump frequency.

The signature of hole burning can be seen even more cleanly in the horizontal cross-sections at three pump frequencies shown in Fig. 2C. At each pump frequency, a bleach band is observed at the frequency approximately corresponding to the pump frequency, although there is a small frequency mismatch due to the overlapping positive signal at the lower frequency side, which we will assign later to the excited-state absorption. These horizontal cross-sections reconfirm that we have indeed observed hole burning. To the best of our knowledge, this is the first experimental observation of the hole-burning effect in the hydrated electron system.

Next, we studied the time evolution of the observed spectral hole by performing the tr-2DES measurements at various waiting times (Fig. 3A). An extremely rapid spectral change is appreciable even in the first 10 fs, and the characteristic diagonally elongated signal at the waiting time of 0 fs is completely lost within 30 fs due to the spectral diffusion, indicating that the system has lost the memory of the pump frequency. In addition to the broadening of the spectral hole, several other spectral features were also observed. Specifically, the 0 fs spectrum shows not only the negative signal due to the ground-state bleach but also a positive signal on the low-frequency side, as we have already mentioned. Furthermore, there is an intense negative signal at the probe frequency of ~12000 $cm^{-1}$, which transiently appears in the spectra measured at the waiting times of 10 and 20 fs.

In general, four types of signals can contribute to tr-2DES spectra, i.e., ground-state bleach, stimulated emission, excited-state absorption, and hot ground-state absorption (*11, 15*). Among these, the hot ground-state absorption appears at a later waiting time when the electronic excitation energy is converted to heat. Therefore, we can neglect its contribution at the waiting time of 0-30 fs. (In Fig. S3 of the Supplementary Materials, the signature of the hot ground state absorption is seen as a positive signal in the low frequency edge of the tr-2DES cross-sections at the waiting time of 500 fs.) The other three signals can appear in early waiting times. Among them, only the excited-state absorption has a positive contribution to tr-2DES spectra. Therefore, the positive signal observed in the low-probe-frequency region of the 0 and 5 fs spectra is attributed to the excited-state absorption of hydrated electrons. Meanwhile, both the ground-state bleach and stimulated emission have negative contributions to tr-2DES spectra. Fortunately, they can be distinguished by their spectral positions. While the ground-state bleach signal is expected to be centered along the diagonal line in tr-2DES spectra, the stimulated emission signal appears only on the lower-probe-frequency side of the diagonal. Accordingly, the negative signal at the probe frequency of ~12000 $cm^{-1}$ in the 10 and 20 fs spectra in Fig. 3A can be attributed to the stimulated emission of hydrated electrons in the excited state. The absence of this signal in the 30 fs spectrum can be interpreted as caused by a rapid downshift of the emission frequency. In fact, a theoretical study reported that the energy gap between the *s*- and *p*-states decreases on a ~25 fs timescale after the photoexcitation of hydrated electrons into a *p*-state (*26*). Finally, the remaining negative signal along the diagonal line can be attributed to ground-state bleach, i.e., hole burning and the subsequent broadening of the spectral hole.

The above spectral assignment indicates that the ground-state bleach signal can be discussed unambiguously by examining the negative signal on the diagonal line or in the lower-right part of the tr-2DES spectra, where the probe frequency is equal to or higher than the pump frequency. To examine the dynamics of hole broadening in more detail, the waiting-time dependence of the tr-2DES spectral intensity was evaluated at the (pump, probe) frequency pairs of (12500 $cm^{-1}$, 12500 $cm^{-1}$) and (12500 $cm^{-1}$, 15000 $cm^{-1}$), as shown by the red curves in Fig. 4. Here, the intensity was integrated over an area of 100 $cm^{-1}$ × 100 $cm^{-1}$ in each tr-2DES spectrum. At the (12500 $cm^{-1}$, 12500 $cm^{-1}$) frequency pair (Fig. 4A, red curve), an intense signal corresponding to the spectral hole in tr-2DES spectra was initially observed, and approximately half of it decayed within 30 fs. In contrast, at the (12500 $cm^{-1}$, 15000 $cm^{-1}$) frequency pair (Fig. 4B, red curve), the bleach signal gradually increased in the first 30 fs. Since this timescale matches the decay of the hole intensity in Fig. 4A, the growth of the bleach signal in Fig. 4B is attributed to the broadening of the hole from the probe frequency of 12500 $cm^{-1}$ to 15000 $cm^{-1}$. The hole broadening arises from the fluctuations of the absorption frequency of each hydrated electron in the ground state. Therefore, this observation indicates that such fluctuations occur on an ultrafast timescale of ~30 fs. This interpretation is supported by a theoretical study that reports ultrafast fluctuation of absorption frequency on a similar timescale, which they attribute to a small translational motion of water molecules that hydrate the electron (*27*). This timescale is also comparable to the sub-50-fs fluctuation of the OH stretch frequency evaluated by a two-dimensional infrared (2DIR) spectroscopy study (*28*), although a more recent 2DIR study reports a considerably longer vibrational frequency correlation time of 176 fs (*29*). On the other hand, the slower decay of the signal in Fig. 4 on the timescale of ~100 fs is attributable to the ground-state recovery following the relaxation from the *p*-state to the *s*-state, which is reported to proceed on the timescale of 60 fs (*24*).

In order to understand the system in further detail, we have also performed tr-2DES measurements by making the pump polarized perpendicular to the probe (Fig. 3B). As mentioned earlier, the observation of the "replica hole" effect is theoretically predicted with this polarization condition, which originates from the asymmetry of the hydrated electrons and the orthogonality of the three transition dipole moments of the $p_x \leftarrow s$, $p_y \leftarrow s$, and $p_z \leftarrow s$ transitions. The comparison between the spectra measured with parallel and perpendicular polarizations in Figs. 3A and B clearly shows the presence of anisotropy at the waiting time of 0 fs, which, however, is quickly lost within 30 fs. This result again indicates very rapid fluctuation of hydrated electrons. Meanwhile, a visual inspection of the spectrum at 0 fs does not reveal any spectral signature for a replica hole, which is expected to appear as a bleach signal in an off-diagonal region of the tr-2DES spectrum measured with perpendicularly polarized pump and probe.

To examine the presence or absence of the replica hole quantitatively, we again evaluated the tr-2DES intensity at the (pump, probe) frequency pairs of (12500 cm$^{-1}$, 12500 cm$^{-1}$) and (12500 cm$^{-1}$, 15000 cm$^{-1}$) as shown in Fig. 4. At the (12500 cm$^{-1}$, 12500 cm$^{-1}$) frequency pair, an intense bleach signal close to time zero is observed only when the polarization of the pump and probe is parallel to each other, implying a strong anisotropy at this (pump, probe) frequency pair. On the other hand, at the (12500 cm$^{-1}$, 15000 cm$^{-1}$) frequency pair, the time traces measured with the two different polarization conditions overlap well, meaning that the anisotropy is not recognized at this (pump, probe) frequency pair. This absence of anisotropy indicates that the replica hole effect is not observed in our experiment. We also performed rigorous analysis that converts the tr-2DES spectra measured with the parallel and perpendicular polarization conditions to diagonal and off-diagonal spectra (*11, 30*), confirming the lack of the replica hole in the observed 2D spectra (Supplementary Text).

The replica hole was predicted based on the assumption that the cavity distortion well splits the three $p \leftarrow s$ absorption bands in energy, such that they appear at different frequencies with bandwidths much narrower than the overall absorption bandwidth of hydrated electrons. Because the absorption spectrum arises from contributions from a large ensemble of hydrated electrons, this assumption implies that a sub-ensemble of hydrated electrons can be selected by using a polarized pump pulse pair that matches the transition energy and transition dipole direction of a particular $p \leftarrow s$ transition (e.g., $p_x \leftarrow s$). The assumption further implies that the three $p \leftarrow s$ transition frequencies are correlated, meaning that the transition frequency distribution of the other two $p \leftarrow s$ transitions (e.g., $p_y \leftarrow s$ and $p_z \leftarrow s$) in this sub-ensemble differs from that of the total ensemble. Consequently, excitation of hydrated electrons into one of the *p*-states is expected to produce a bleach signal at a different frequency corresponding to the other two $p \leftarrow s$ transitions when probed with the pulse polarized perpendicular to the pump. Our experimental results, however, indicate that the frequency distribution of these other two $p \leftarrow s$ transitions in the sub-ensemble is indistinguishable from that of the total ensemble. This implies the lack of correlation among the three $p \leftarrow s$ transition frequencies, which explains why hole burning is observed for parallel pump-probe polarization but not for perpendicular polarization. The absence of the frequency correlation is considered a natural consequence of a large distribution of shapes and sizes of the electron-trapping cavity in liquid water. In other words, our experimental results indicate that hydrated electrons are highly structurally inhomogeneous, trapped in broadly distributed and nonuniform cavities, which is consistent with our notion that liquid water is a highly flexible system.

Finally, we discuss the reason why we were able to observe the spectral hole-burning effect, while it was not detected in previous experimental attempts. The main difference lies in the time resolution of the optical measurements. In the pump-probe measurements done by the Laubereau and Schwartz groups, the time resolution was 100-200 fs (*13, 14*). Considering that the broadening of the hole occurs within 30 fs as we see in Fig. 3, their time resolution was not sufficient for observing the hole that survives only transiently. Meanwhile, the Wiersma group used 5 fs pulses in their measurements, and the time resolution was sufficient to resolve the short-lived holes. In their study, however, only the spectra with the pump-probe delays at and later than 20 fs are presented (*15*). This is most likely the reason why the hole was not observed in their study. This brief review of previous studies shows that the present study reports the first experiments that provide the hole burning spectra with a delay shorter than 20 fs with sufficient time resolution. We note that our data are mostly free from coherent artifacts that often appear when multiple pulses used in the measurements are temporally overlapped (see the Material and Methods section of the Supplementary Materials). The high time resolution combined with the careful elimination of the coherent artifacts enabled us to observe the hole-burning effect in the hydrated electron system.

In summary, using tr-2DES, we have experimentally shown that hydrated electrons have a structural inhomogeneity, with their shape fluctuating on a timescale shorter than 30 fs. This result reveals the unique nature of electron confinement within a flexible "box" in liquid water, which is distinct from the electron confinement in rigid solid-state systems.

**Acknowledgments:** We thank Mr. Masaharu Watanuki and his colleagues at Advanced Manufacturing Support Team, RIKEN Center for Advanced Photonics (RAP) for the construction of the wire-guided liquid jet system and the retroreflector.

**Funding:**

JST PRESTO Grant Number JPMJPR17P4 (HK)

JSPS KAKENHI Grant Numbers JP21K18943, JP25K22248 (TT)

**Author contributions:**

Conceptualization: HK, TT

Methodology: KM, HK, TT

Investigation: KM, HK

Funding acquisition: HK, TT

Supervision: TT

Writing – original draft: KM, TT

Writing – review & editing: KM, HK, TT

**Competing interests:** Authors declare that they have no competing interests.

**Data and materials availability:** All data are available in the main text or the supplementary materials.

**Supplementary Materials**

Materials and Methods

Supplementary Text

Figs. S1 to S4

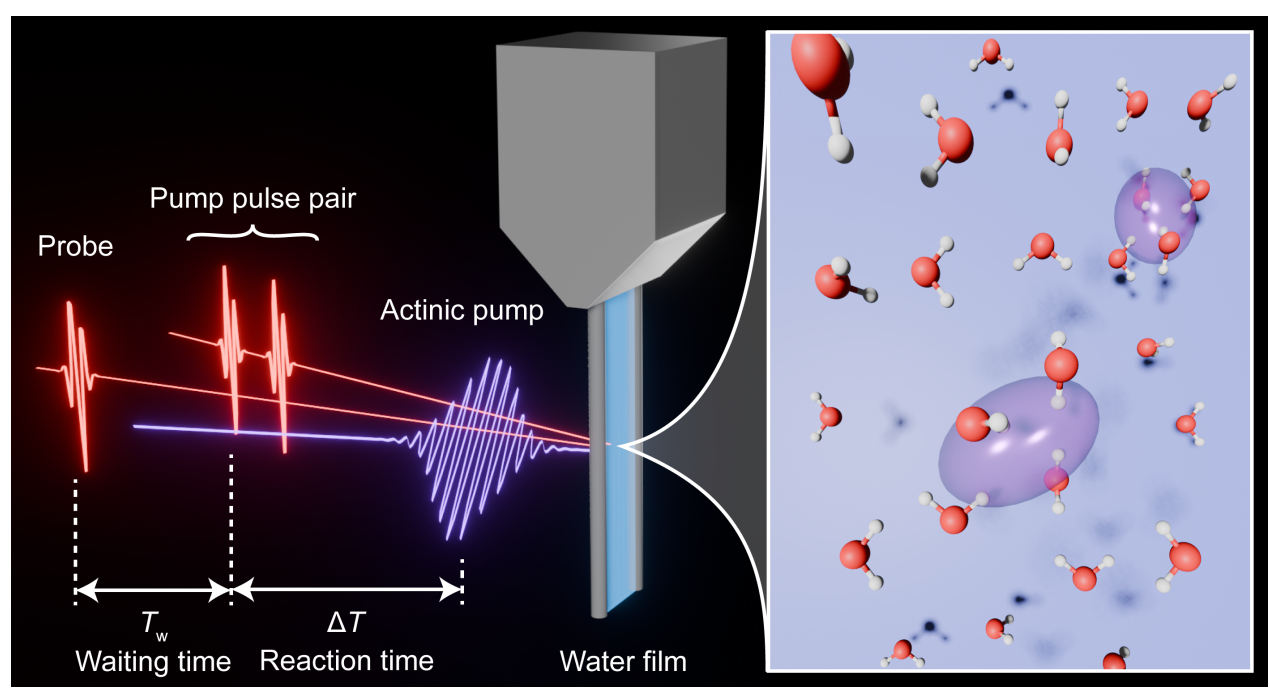

**Fig. 1. Schematic of tr-2DES measurements of hydrated electrons.** An actinic pump, a pump pulse pair, and a probe were sequentially irradiated onto a thin film of liquid water to perform the measurements. The inset shows a conceptual drawing of hydrated electrons in liquid water, where the transparent purple ellipsoids represent electrons.

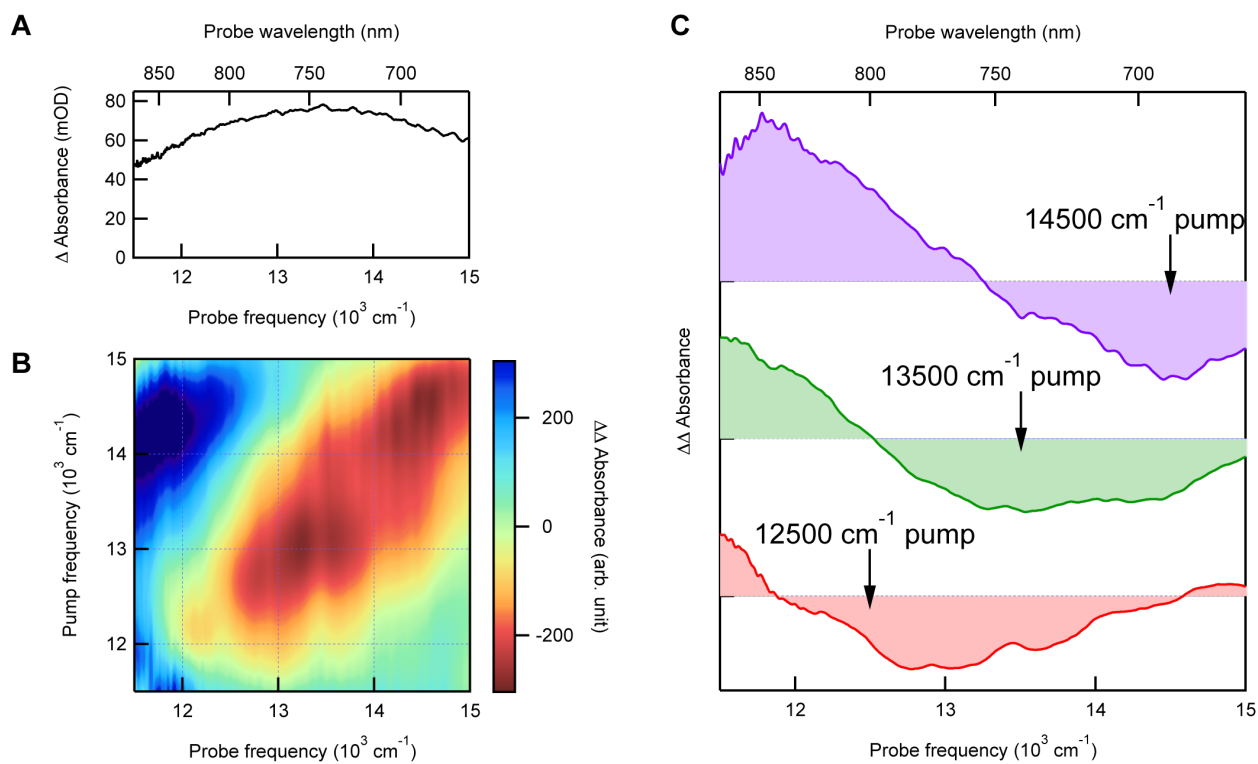

**Fig. 2. Tr-2DES of hydrated electrons.** (**A, B**) A transient absorption spectrum (**A**) and a tr-2DES spectrum (**B**) of transiently generated hydrated electrons. Electrons were introduced into water 10 ps prior to these measurements. The waiting time in the tr-2DES measurement was set to 0 fs. (**C**) Horizontal cross sections of the tr-2DES spectrum in **B** at the pump frequencies of 14500, 13500, and 12500 cm$^{-1}$ (from top to bottom).

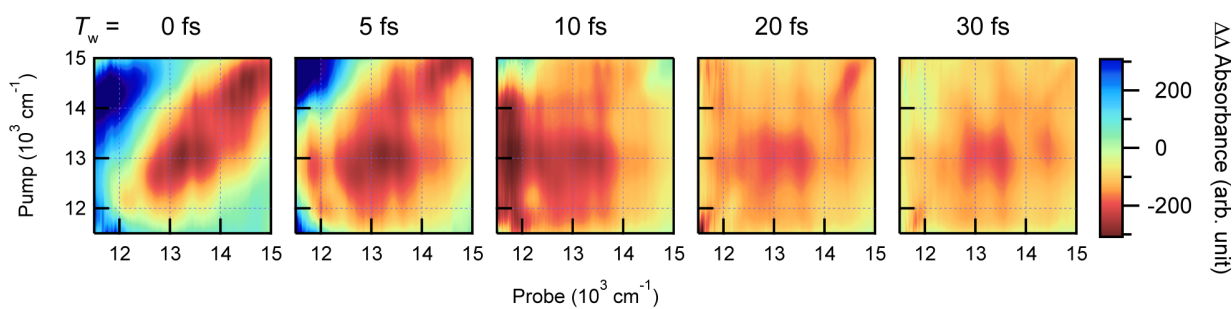

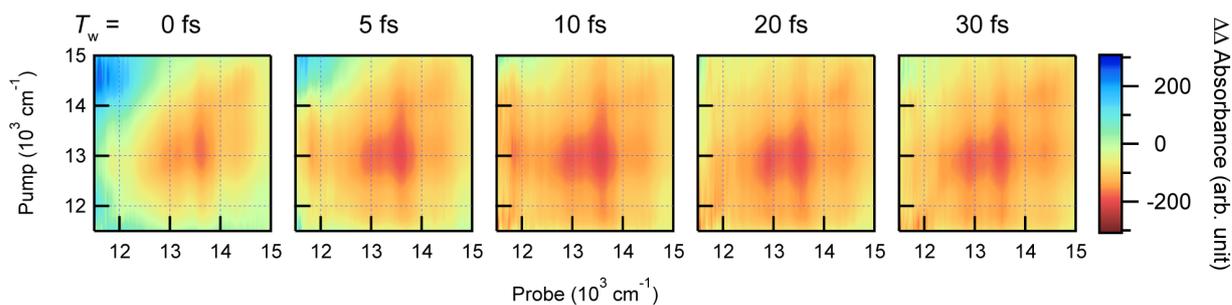

**Fig. 3. Waiting time and polarization dependence of the tr-2DES spectra of hydrated electrons.** The pump pulse pair was polarized parallel (**A**) or perpendicular (**B**) with respect to the probe.

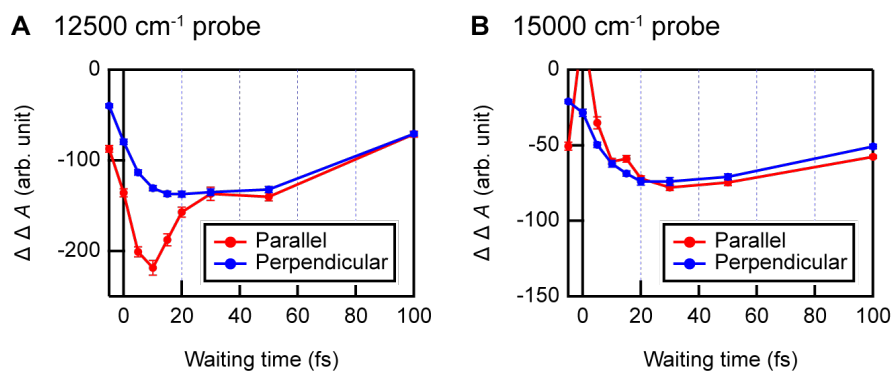

**Fig. 4. Tr-2DES spectral intensity of hydrated electrons as a function of waiting time and polarization.** The intensity was evaluated at the pump frequency of 12500 cm$^{-1}$ and the probe frequency of 12500 cm$^{-1}$ (**A**) or 15000 cm$^{-1}$ (**B**). The polarization of the pump pulse pair and the probe was parallel for the red curves and perpendicular for the blue curves. The error bars indicate the 68% confidence interval.

# Supplementary Materials for

# Electron confinement within a fluctuating "box" in liquid water

Korenobu Matsuzaki, Hikaru Kuramochi, and Tahei Tahara

Corresponding author: tahei@riken.jp

**The PDF file includes:**

    Materials and Methods
    Supplementary Text
    Figs. S1 to S4

**Materials and Methods**

Materials

High-purity water (18.2 MΩ·cm resistivity) was obtained from a commercial water purification system (Millipore, Milli-Q Advantage A10). A thin film of this high-purity water was prepared by a homebuilt wire-guided liquid jet system (*31*) and was used as the sample. In this system, the high-purity water stored in a reservoir in the upper part of the system smoothly flowed down along the two stainless wires, producing a thin water film between the two wires with an approximate thickness of 400 μm. After flowing down, the water was pumped back to the upper reservoir using a gear pump (MicroPump, GAF-T23.DGF61.J).

Method: Transient two-dimensional electronic spectroscopy (tr-2DES)

A Yb:KGW regenerative amplifier system (Light Conversion, PHAROS-SP; 1030 nm, 190 fs, 1 mJ, 1 kHz) was used as the light source for preparing an actinic pump pulse, a pump pulse pair, and a probe pulse that are required to perform transient two-dimensional electronic spectroscopy (tr-2DES) measurements (Fig. S1) (*32, 33*).

The actinic pump pulse at 257 nm was obtained as the fourth harmonic of the amplifier output. To this end, 0.5 mJ of the amplifier output at 1030 nm was converted to the second harmonic at 515 nm using β-barium borate (β-BBO; θ = 27.7°, thickness = 1.5 mm), which was further converted to the fourth harmonic using another β-BBO (θ = 50.1°, thickness = 5 mm). The delay of the actinic pump pulse ("reaction time", $\Delta T$, Figs. 1 and S1) was set to a desired value using an optical delay stage equipped with a homebuilt retroreflector. Subsequently, its polarization was adjusted to the magic angle (54.7°) with respect to the probe, and it was focused onto the sample using a planoconvex lens (f = 250 mm).

The pump pulse pair and the probe were obtained from a homebuilt noncollinear optical parametric amplifier (NOPA), which was powered by 0.4 mJ of the amplifier output. A small portion of the amplifier output was converted to white light by self-phase modulation in a sapphire plate (thickness = 3 mm). After rejecting the long wavelength component above 960 nm using a short pass filter, this white light was used as a seed for the NOPA. The remainder of the amplifier output was converted to the second harmonic at 515 nm using β-BBO (θ = 23.4°, thickness = 2 mm) and was used as the pump for the NOPA. By temporally and spatially overlapping the seed and the pump inside β-BBO (θ = 24.5°, thickness = 1.5 mm), broadband visible light (620-910 nm) was obtained as the NOPA output. The chirp compensation of the generated pulse was done using a micromachined membrane deformable mirror (MMDM) (*34*). First, the pulse was diffracted by a grating, and each spectral component was focused onto different positions of a 19-channel MMDM (OKO Tech) using a concave mirror. In this way, each spectral component can acquire different phase offsets depending on the shape of the MMDM at each position. Each of the spectral components reflected by the MMDM was collimated by the same concave mirror and was combined again into a single beam using the same grating. The shape of the deformable mirror was optimized using a genetic algorithm so as to maximize the intensity of the sum frequency of the pump pulse pair (with the two pulses temporally overlapped) and the probe, which was generated with β-BBO (θ = 36.7°, thickness = 8 μm) at the sample position.

A major part of this NOPA output was converted to a pump pulse pair using the translating-wedge-based identical pulses encoding system (TWINS; inset in Fig. S1) (*35*). The chirp introduced within TWINS was compensated by using a pair of chirped mirrors (Laser Quantum, DCM9) and by adjusting the insertion length of the Z-cut α-BBO wedge in the second wedge pair in TWINS. The generated pump pulse pair was separated into two using a beam splitter (R:T = 4:1). For the reflected portion, the polarization was set to horizontal or vertical using a half-wave plate, and it was focused onto the sample using a concave mirror (r = 500 mm). In order to suppress the artifact in tr-2DES spectra due to the interference between the probe and the scattered pump pulse pair, the delay of the pump pulse pair was continuously modulated using a piezo stage (*36*). On the other hand, the portion of the pump pulse pair that was transmitted through the beam splitter went through a sensitivity correction filter (Thorlabs, SRF11), and it was detected by a Si photodiode (Hamamatsu, S2281). By measuring the pump pulse pair intensity as a function of the temporal separation between the paired pulses ("coherence time", Fig. S1), we were able to deduce the spectrum of the pump pulse pair based on the Fourier transformation analysis (Fig. S2C).

The remaining portion of the NOPA output was separated into the probe and the reference using a beam splitter (R:T = 1:1). The probe was delayed by a desired amount using an optical delay stage with respect to the second pulse of the pump pulse pair ("waiting time", $T_w$, Figs. 1 and S1), and was focused alongside the pump pulse pair onto the sample using a concave mirror (r = 500 mm). After the transmission through the sample, the probe was sent to a polychromator (Horiba Jobin Yvon, iHR320) using an optical fiber, and it was eventually detected by a photodiode array (Hamamatsu, S3904-512Q). The reference pulse was also introduced into the same polychromator using another optical fiber, and it was detected simultaneously with the probe using another photodiode array (Hamamatsu, S3904-512Q).

The temporal resolution along the waiting time axis was evaluated by measuring the cross-correlation of the probe and the pump pulse pair (the paired pulses were completely overlapped in time for this measurement). To do this, β-BBO (θ = 36.7°, thickness = 8 μm) was placed at the sample position, and the sum frequency intensity of the two pulses was measured as a function of the waiting time. The cross-correlation width of the two pulses determined in this manner was 12.5 fs FWHM (full width at half maximum). If we assume that the two pulses have a Gaussian temporal profile with the same width, this result indicates that the FWHM of the probe and the pump pulse pair was 9 fs. In a similar manner, the temporal resolution along the reaction time axis was evaluated by irradiating the probe and the actinic pump onto the same β-BBO (θ = 36.7°, thickness = 8 μm), and by detecting the difference frequency between the two. The cross-correlation width obtained in this manner was 370 fs FWHM.

In order to obtain 2DES spectra, we need to evaluate the change in the probe spectrum caused by the pump pulse pair. To realize this, a chopper was used to chop the pump pulse pair at 0.5 kHz so that we could measure the probe spectrum with and without the pump pulse pair irradiation ($P$(pump pulse pair) and $P_0$). The 2DES signal ($S_0$) was obtained by detecting the following difference.

$$S_0 = P(\text{pump pulse pair}) - P_0 \qquad (S1)$$

In our study, we measured tr-2DES, which corresponds to the difference between the 2DES signals with and without the actinic pump ($S$(actinic) and $S_0$). In order to detect this signal, we used another chopper to chop the actinic pump at 0.25 kHz. Using these two choppers, the probe spectra were recorded under four different conditions (with/without the actinic pump and with/without the pump pulse pair). The tr-2DES signal was then extracted in the following manner.

$$S(\text{actinic}) - S_0 = [P(\text{actinic pump + pump pulse pair}) - P(\text{actinic pump})] \\ - [P(\text{pump pulse pair}) - P_0] \qquad (S2)$$

This measurement and analysis procedure provides us with a robust method for obtaining reliable tr-2DES spectra even when the waiting time is close to 0 fs. At such waiting times, one of the pulses in the pump pulse pair is temporally overlapped with the probe pulse, and the obtained 2DES spectra are often contaminated by the so-called coherent artifacts. The most common type of coherent artifacts arises from the nonresonant interaction with the sample, and it is expected that we observe the same coherent artifacts irrespective of the presence or absence of the actinic pump. In other words, $S$(actinic) and $S_0$ in the equation above contain the same coherent artifacts. Therefore, by subtracting $S_0$ from $S$(actinic), it is possible to eliminate the coherent artifact and isolate the true tr-2DES signal in a clean manner. This allowed us to obtain reliable tr-2DES spectra even at the waiting time of 0 fs. In addition to those nonresonant coherent artifacts, the resonant interaction with the sample gives rise to another type of coherent artifacts, such as perturbed free induction decay (PFID). Although the resonant coherent artifacts cannot be eliminated by the procedure mentioned above, the severe spectral contamination by PFID occurs in the negative waiting time, and its contribution is not significant at time zero and in the positive waiting time (*37*).

The experimental data obtained in the aforementioned manner were analyzed in the same way as in the literature (*36*) to yield tr-2DES spectra. The tr-2DES spectra constructed in this manner are distorted along the pump axis due to the frequency dependence of the pump pulse pair intensity (Fig. S2). In order to correct this spectral distortion, the tr-2DES spectra shown in this paper (except for Fig. S2) were normalized along the pump axis using the pump pulse pair spectrum.

Method: Transient absorption spectroscopy

Transient absorption spectra were measured using the tr-2DES setup described in detail above by simply blocking the pump pulse pair.

**Supplementary Text**

Conversion of the tr-2DES spectra to diagonal and off-diagonal spectra

As explained in the main text, the main part of the absorption spectrum of hydrated electrons is attributed to the three $p \leftarrow s$ transitions with orthogonal transition dipole moments. By taking advantage of the orthogonality of those three transition dipole moments, in Fig. 3, we address each transition selectively by controlling the polarization of the pump pulse pair and probe. In particular, in the experiment with the perpendicular polarization condition, we aimed to, e.g., pump the electron using the $p_x \leftarrow s$ transition and probe the $p_y \leftarrow s$ and $p_z \leftarrow s$ transitions. In reality, however, the situation is more complicated because hydrated electrons are randomly oriented and their transition dipole moments are not perfectly parallel or perpendicular to the polarization direction of the pump pulse pair and probe. As a consequence, even when the pump pulse pair and probe are polarized perpendicular to each other, there is a possibility that the pump pulse pair and probe interact with the same transition when the relevant transition dipole is tilted with respect to the polarizations of the pump pulse pair and the probe pulse. In fact, the bleach lobe in the spectrum at $T_w = 0$ fs in Fib. 3B measured with the perpendicular polarization condition is slightly elongated along the diagonal because of this reason. This complication can be resolved with a mathematical procedure that provides diagonal and off-diagonal spectra (*11, 30*).

For a system with triply degenerate excited states as in the case of hydrated electrons, the experimentally obtained tr-2DES spectra with the pump pulse pair polarized parallel or perpendicular to the probe ($\Delta\Delta A_\parallel$ and $\Delta\Delta A_\perp$) can be converted to diagonal and off-diagonal spectra ($\Delta\Delta A_{\text{diag}}$ and $\Delta\Delta A_{\text{off-diag}}$) given by the equations below (*11, 30*).

$$\Delta\Delta A_{\text{diag}} = \frac{2}{3}\Delta\Delta A_\parallel - \frac{1}{3}\Delta\Delta A_\perp \tag{S3}$$

$$\Delta\Delta A_{\text{off-diag}} = -\frac{1}{6}\Delta\Delta A_\parallel + \frac{1}{2}\Delta\Delta A_\perp \tag{S4}$$

The diagonal spectrum, $\Delta\Delta A_{\text{diag}}$, captures only the contribution in which pump and probe interact with the same transition. In the meantime, the off-diagonal spectrum, $\Delta\Delta A_{\text{off-diag}}$, involves only the processes where pump and probe interrogate different transitions with orthogonal transition dipole moments. Therefore, the diagonal and off-diagonal spectra enable us to fully exploit the orthogonality of the three transition dipole moments in hydrated electrons, allowing us to discuss, for example, the presence or absence of the replica hole effect in a clean manner.

Fig. S4 shows the diagonal and off-diagonal spectra obtained by converting the spectra in Fig. 3 using the equations above. The diagonal spectrum at $T_w = 0$ fs shows a diagonally elongated signal, which is similar to the one in Fig. 3A measured under the parallel polarization condition. On the other hand, the off-diagonal spectrum at $T_w = 0$ fs exhibits a uniform spectral distribution, which is distinct from the spectrum in Fig. 3B measured under the perpendicular polarization condition. The uniform spectrum obtained in this manner endorses our conclusion in the main text that the replica hole effect is not observed in our data.

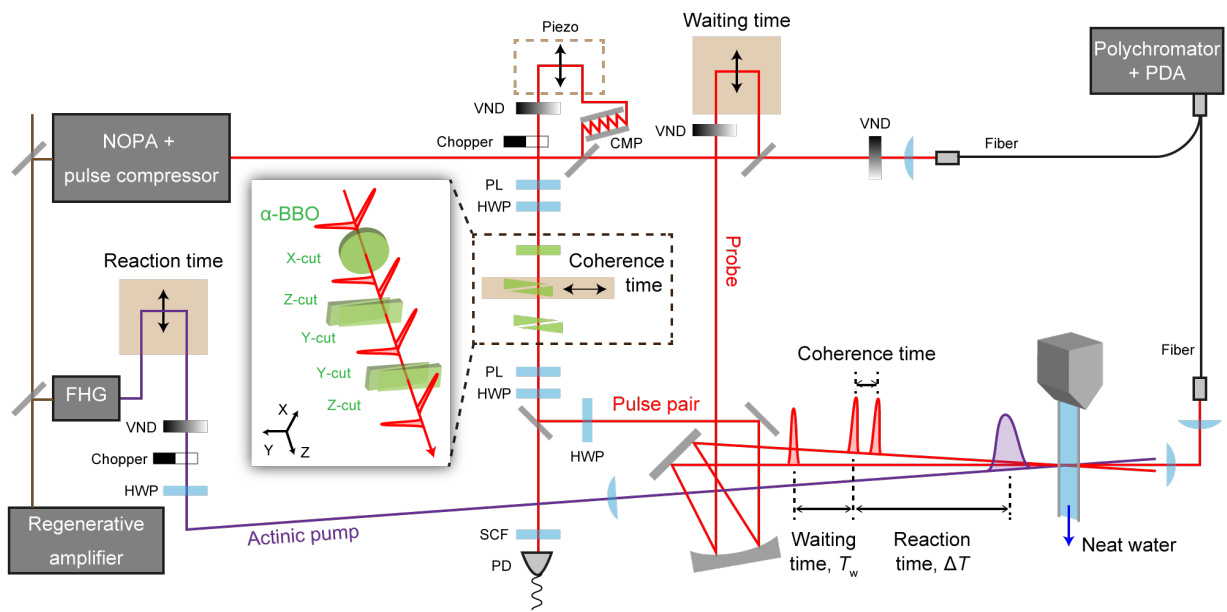

**Fig. S1. Schematic of the experimental setup.** Transient two-dimensional electronic spectroscopy (tr-2DES) setup used in this study. (BBO: barium borate, CMP: chirped mirror pair, FHG: fourth harmonic generation, HWP: half-wave plate, NOPA: Noncollinear optical parametric amplifier, PD: photodiode, PDA: photodiode array, PL: polarizer, SCF: sensitivity correction filter, VND: variable neutral density filter)

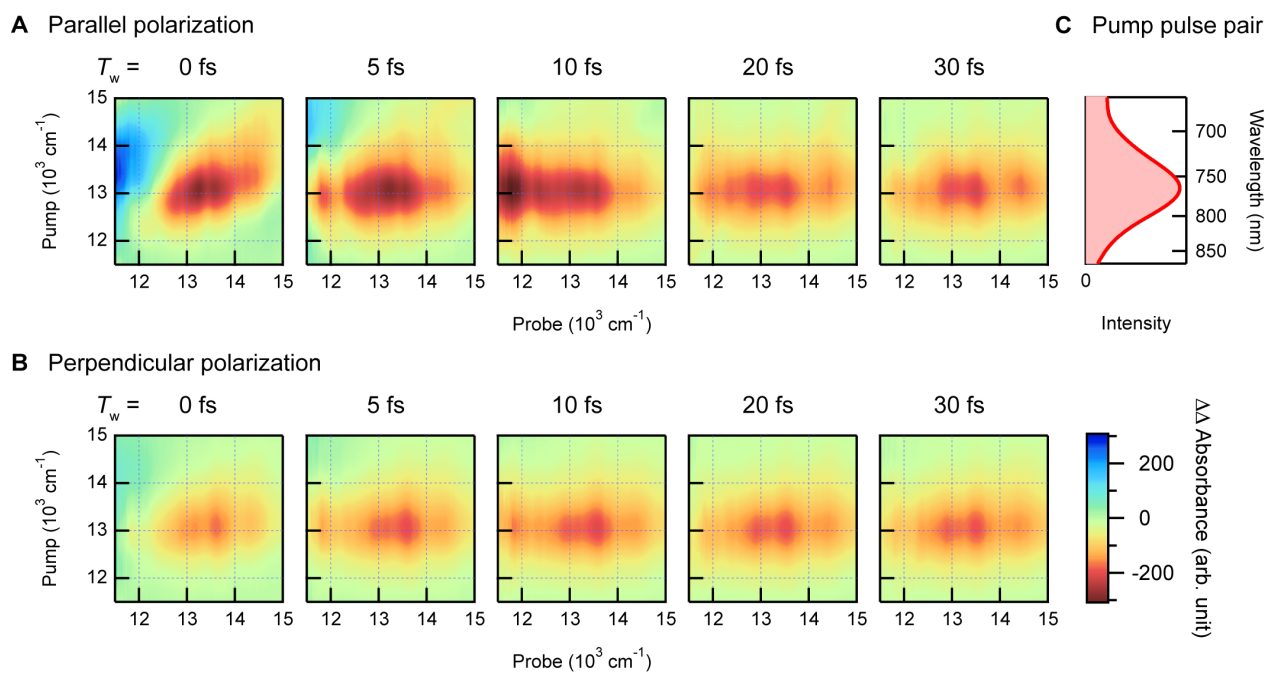

**Fig. S2. Tr-2DES spectra of hydrated electrons without normalization by the pump pulse pair spectrum.** Spectra at each waiting time $T_w$ were measured with the pump pulse pair polarized parallel (**A**) or perpendicular (**B**) to the probe. Normalizing these spectra using the pump pulse pair spectrum (**C**) yields the tr-2DES spectra in Fig. 3.

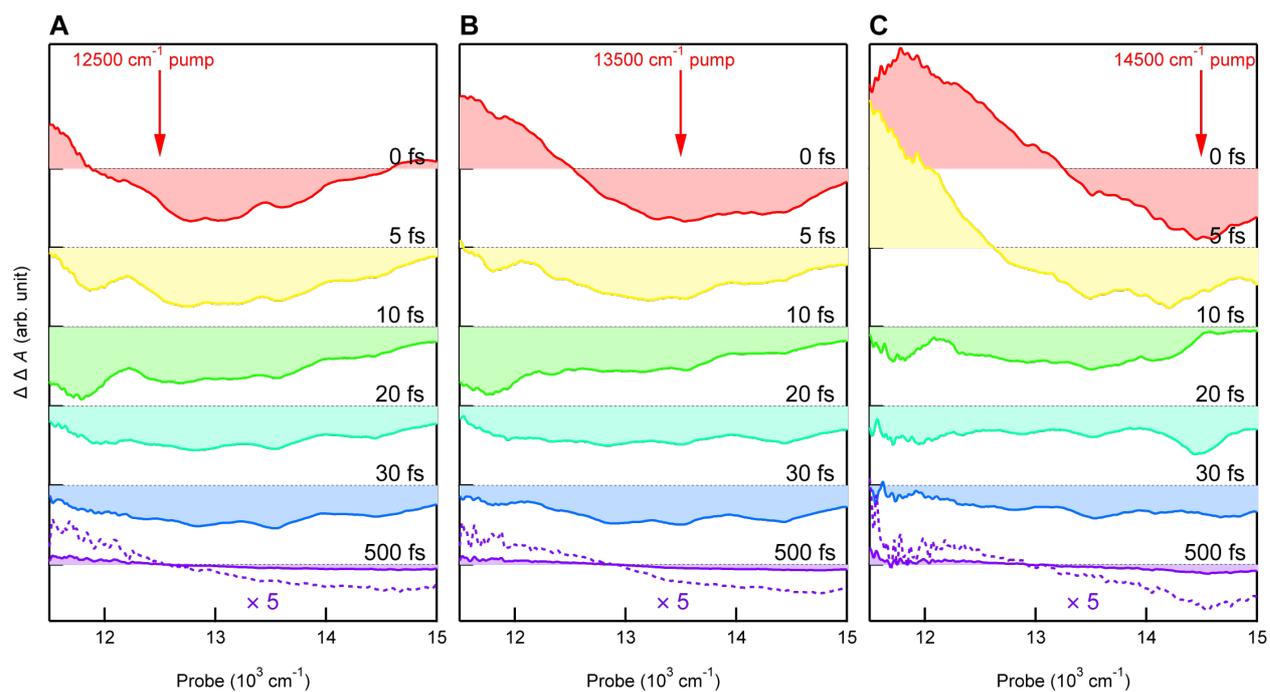

**Fig. S3. Horizontal cross-sections of tr-2DES spectra of hydrated electrons.** The cross-sections at each waiting time were evaluated at the pump frequencies of 12500 cm$^{-1}$ (**A**), 13500 cm$^{-1}$ (**B**), and 14500 cm$^{-1}$ (**C**).

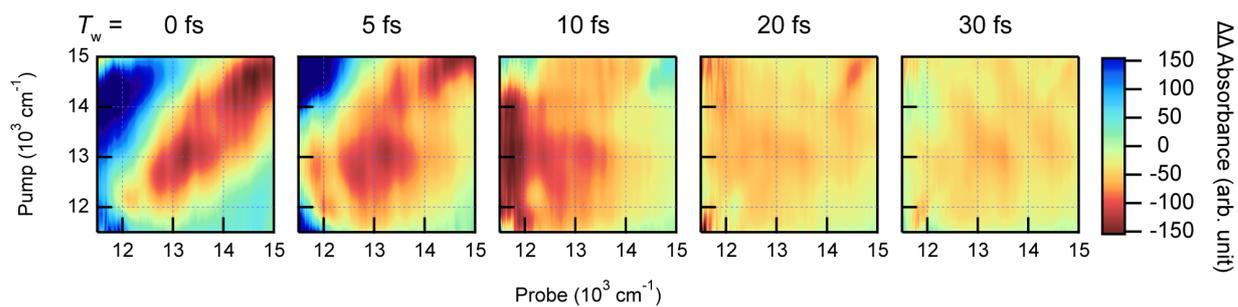

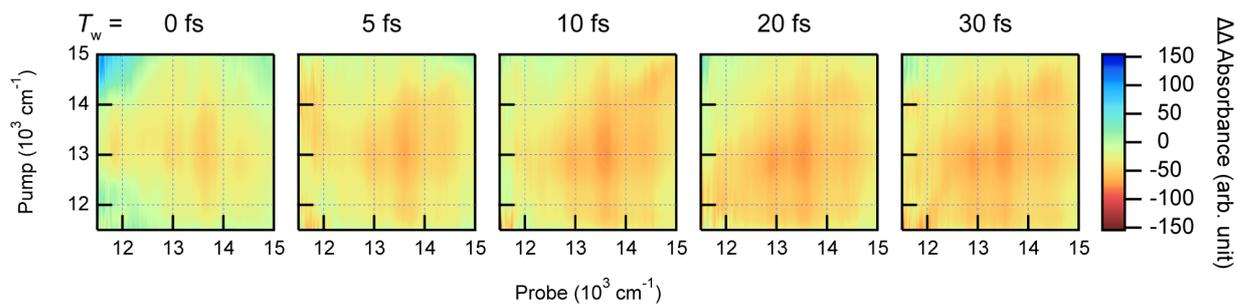

**Fig. S4. Diagonal and off-diagonal tr-2DES spectra of hydrated electrons.** The tr-2DES spectra in Fig. 3 measured under the parallel and perpendicular polarization conditions were converted to diagonal and off-diagonal spectra using Eqs. (S3) and (S4).